\documentclass[aps,prb, twocolumn]{revtex4}
\usepackage{graphicx}
\usepackage{psfrag}
\usepackage{subfigure}
\usepackage[dvips]{color}

\usepackage{times}
\usepackage{pslatex}
\usepackage{amsmath}
\usepackage{amsfonts}
\usepackage{amsthm}
\usepackage{amssymb}
\usepackage{amsbsy}
\usepackage{pstricks}
\usepackage{wasysym}
\usepackage{mathrsfs}
\usepackage{bbm}
\usepackage{stmaryrd}

\input{epsf}

\newcommand{\beq}{\begin{equation}}
\newcommand{\eeq}{\end{equation}}
\newcommand{\bea}{\begin{eqnarray}}
\newcommand{\eea}{\end{eqnarray}}

\begin{document}

\draft
\title{Diagnosing Deconfinement and Topological Order}
\author {K. Gregor$^1$, David A. Huse$^2$, R. Moessner$^3$, S. L. Sondhi$^2$}
\address{$^1$Courant Institute, New York University, New York, NY, 10003\\
$^2$Department of Physics, Princeton University, Princeton, NJ 08544\\
$^3$Max Planck Institute for the Physics of Complex Systems, 01187 Dresden}
\date{\today}
\begin{abstract}
Topological or deconfined phases are characterized by emergent, weakly
fluctuating, gauge fields. In condensed matter settings they inevitably
come coupled to excitations that carry the corresponding gauge charges
which invalidate the standard diagnostic of deconfinement---the Wilson
loop. Inspired by a mapping between symmetric sponges and the deconfined
phase of the $Z_2$ gauge theory, we construct a diagnostic for deconfinement
that has the interpretation of a line tension. One operator version of this
diagnostic turns out to be the Fredenhagen-Marcu order parameter known to
lattice gauge theorists and we show that a different version is best suited
to condensed matter systems. We discuss generalizations of the diagnostic,
use it to establish the existence of finite temperature topological
phases in $d \ge 3$ dimensions and show that multiplets of the diagnostic are
useful in settings with multiple phases such as $U(1)$ gauge theories with
charge $q$ matter. [Additionally we present an exact reduction of the partition
function of the toric code in general dimensions to a well studied problem.]

\end{abstract}
\maketitle

\section{Introduction}
\label{sec:intro}

There is currently much interest in condensed matter systems that exhibit
ordering captured by a gauge theory lacking a local order parameter. Such
systems offer a contrast to the classical broken symmetry paradigm built
around the notion of an order parameter introduced by Landau and these phases said to be 
topological or deconfined\cite{wenbook,RMPtopo} as they also exhibit quasiparticles
with fractional quantum numbers \cite{Rajaraman}. 
These phases are of interest in the study
of strongly correlated quantum systems---where the quantum Hall states \cite{wenniu}
and resonating valence bond liquids \cite{pwaRVB,RMtriRVB} are the canonical examples. Their
interest has been further enhanced by Kitaev \cite{KitTopo} and Freedman's \cite{RMPtopo} 
proposal of utilizing their subset possessed of non-abelian braiding statistics for
quasiparticles, for the construction of a quantum computer.

The theoretical description of such phases is in terms of a gauge theory, possibly with
low energy matter.  In the simplest cases the low-energy theory is a purely topological gauge theory such as Chern-Simons theory in the case of the quantum Hall states \cite{wenniu} or the 
BF theory in the case of resonating valence bond liquids \cite{HOSbf}. Indeed, the term topological phase dates from these early instances; absent a better standard term, we will use it here to refer to all phases with an emergent gauge  field. The explosion of recent interest has come from the construction of a wide variety of lattice models that realize a variety of long wavelength theories including those with dynamical gauge fields \cite{wenbook,HenleyRev}.

In idealized models the existence of such phases is transparent for one can readily show that
the gauge field: (a) exhibits weak fluctuations and thus a perimeter law for the Wilson loop in ground states, (b) mediates a non-confining force between its sources (``quarks''), and
(c) [where the low energy theory is purely topological] leads to a topology dependent
ground state degeneracy.
For more realistic Hamiltonians, unfortunately, this transparency is lost. The problem with
the first two characterizations on our list is that, as is well known from the lattice gauge
theory literature, they fail in the presence of dynamical sources (``matter'') for the gauge field.

In the condensed matter setting, the gauge field is only emergent and {\it necessarily} comes
with higher energy excitations that carry a  gauge charge, making this a generically fatal flaw.
The situation only gets worse at finite temperatures where such matter excitations must enter the statistical sum. The third characterization can continue to work in strictly topological phases precisely at $T=0$ but lacks a clear meaning beyond that limit and outside that subset of
deconfined phases.\footnote{There {\it is} a spectral test---the deconfined phase has particles
that carry the charge of the deconfined gauge field and also characteristically exhibit
fractionalization of the microscopic quantum numbers. For example, in our workhorse example later in this paper we can ask if the Hamiltonian leads to isolable particles with $Z_2$ [\onlinecite{FradkinShenker}]. However, this test requires a fair amount of information and
in the presence of bound states of the fractionalized objects even more so. Note also
that fractionalized quantum numbers are not quantized signatures of a deconfined phase
\cite{IrrationalCharge}.}

All of this is an unsatisfactory state of affairs. As in Landau theory, it would be nice to
have a fixed time diagnostic operator that one can compute when handed a ground state or a Gibbs
state to decide whether it exhibits the ordering characteristic 
of a given topological phase.\footnote{Of the existing proposals for diagnosing 
topological phases, the one which closest in spirit to such a prescription is the 
computation of an entanglement entropy, 
using a combination of appropriately defined reduced density matrices for a system 
subjected to different partitionings: 
A. Kitaev, J. Preskill, Phys.\ Rev.\ Lett.\ {\bf 96}, 110404 (2006); M. Levin and X.-G. Wen, {\it ibid.} 110405.} In
this paper we report the construction of such a diagnostic which teases out the underlying weakly fluctuating behavior of the gauge
field. We do so  via a detour into the theory of sponge phases where our diagnostic has the interpretation of a line tension---picking up a line of analysis begun by Huse and
Leibler \cite{HuseLeibler} a while back. Remarkably, this diagnostic turns out to be a spacetime
generalization of the so-called Fredenhagen-Marcu order parameter \cite{FMop1,FMop2} known to 
lattice gauge theorists.

In this paper we also begin the process of applying these ideas to a wide class of systems. While
the bulk of our paper is concerned with the $Z_2$ gauge theory with matter (or Kitaev's toric code
\cite{KitTopo} with generic perturbations) we sketch the generalization to different low energy
gauge structures
including cases where it is necessary to use a multiplet of diagnostics. Notably, we use the diagnostic to establish the survival of topological phases to finite temperatures in $d \ge 3$\cite{SentFish3d}.

Turning to the contents of the paper, we begin in Sections II and III by reviewing the formulation of the $Z_2$ gauge theory with $Z_2$ matter and its reformulation as a statistical mechanics of surfaces with edges. This leads us to the identification of the line tension diagnostic in Section
IV and its operator formulations in Section V. In Section VI we discuss why condensed matter
systems prefer a particular operator formulation. Sections VII and VIII deal with applications
of the ideas to finite temperature phases and more complex phase diagrams. The conclusion flags some
open questions. Finally, two appendices establish some useful results on Kitaev's toric code: firstly, we obtain its partition function in general dimension via a reduction to a classical pure $Z_2$ gauge theory; secondly, we demonstrate perturbative stability of its topological order \cite{nR1,nR2}, and find it to persist to finite temperature only for $d \geq 3$, in accordance with Ref.~[\onlinecite{Claudios3d}].

Before proceeding, we should point out an important antecedent to our work which also addressed
the challenge of exhibiting topological order away from the idealized models. Hastings and Wen 
\cite{HastWen} gave
a $T=0$ continuity construction starting with idealized Hamiltonians that generates a {\it Hamiltonian--dependent} gauge field operator  which exhibits a perfect perimeter law (with zero coefficient) for the Wilson loop in deconfined phases. By contrast, our construction is Hamiltonian--independent in the spirit of an order parameter.

\section{$Z_2$ Gauge Theory With Matter}
\label{sec:z2gt}

Consider the Hamiltonian of the $Z_2$ lattice
gauge theory,
\beq
\label{IGTHamil}
-H = K \sum_p \prod_{l \in \partial p} \sigma^z_l +
 \Gamma \sum_l \sigma^x_l
+ J \sum_l \sigma^z_l \prod_{s \in \partial l} \tau^z_s
+ \Gamma_M \sum_s \tau^x_s
\eeq
supplemented by the constraint that we restrict its action
to ``gauge invariant'' states defined by
\begin{equation}
G_s |\psi \rangle = | \psi \rangle \ \ \ , \ \ G_s= \tau^x_s
\prod_{l: s \in \partial l} \sigma^x_l \ ,
\label{eq:IGTconstraint}
\end{equation}
where the gauge ($\sigma^i_l$) and  matter ($\tau^i_s$) operators act in spin $1/2$ Hilbert spaces that live on the links $l$ and sites $s$, respectively, of a (hyper) cubic lattice in $d$ dimensions. The subscripts
$s$, $l$ and $p$ denote sites, links and plaquettes. $\partial p$ and $\partial l$
are the boundaries of the corresponding objects.

For $\Gamma=J=0$ our Hamiltonian is equivalent to Kitaev's toric code as noted in the
original paper.\cite{KitTopo} 
In that case the degenerate $2^{N_{nc}}$ ground states ($N_{nc}$ is the number of
independent non-contractible loops on the lattice) are easily constructed and exhibit a
perfect ``perimeter'' law for the contractible Wilson loop
\beq
W[C] =  \langle \prod_{l \in C} \sigma^z_l \rangle = 1
\eeq
Consistent with this ground state description, one sees that the excited states
contain non-interacting charges, free to sit at arbitrary locations in this particular
model. The remaining part of the excitation spectrum is described by purely gauge excitations.
These are vortices of the gauge field, or visons, which are isolated plaquettes
for which $W[C]=-1$. Both charged and gauge excitations are separated from the set
of ground states by a finite gap.

The full $T=0$ phase diagram of the Hamiltonian (\ref{IGTHamil}) has two distinct phases.
The first is the deconfined phase that exists for small perturbations of the Kitaev point. (For a more detailed analysis of the toric code phase diagram, including $T > 0$, see App. \ref{app:kita}.) The remaining phase is the confined-Higgs phase which takes its name from the limiting regions that were shown to be smoothly connected by Fradkin and Shenker \cite{FradkinShenker}.

How do we distinguish between the two phases? Unfortunately, except along
the line $J/\Gamma_M=0$, the Wilson loop $W[C]$ does not readily detect
the phase transition---it exhibits a perimeter law {\it everywhere} (and also
at $T > 0$). As previously remarked, the potential between two test charges
is also not a sharp diagnostic due to the presence of dynamical matter.

\section{Surfaces, Edges and the Symmetric Sponge}
\label{sec:spongemapping}

In order to get to our diagnostic, let us now reformulate the Hamiltonian gauge theory considered
above as the classical statistical mechanics of a system of membranes. To do this we first write
down the path integral formulation with a discretized imaginary time. This is governed by the
action \cite{kogutRMP}:
\bea
\label{eq:z2action}
-S=K \sum_p \prod_{l \in \partial p} \sigma_l +J \sum_l \sigma_l \prod_{s \in \partial l} \tau_s
\ ,
\eea
where the $\sigma_l$ are the gauge fields living on the links, and $\tau_s$ the matter fields
residing on the sites of a $d+1$ dimensional hypercubic lattice, with $d \geq 2$ so
that we can discuss a deconfined phase.
In writing this spacetime symmetric form, we have moved away from the anisotropic limit that is
needed for precise reproduction of the Hamiltonian problem but which is not important
for our purposes. The reader should also bear in mind that $K$ and $J$ refer to different quantities, with
different units, in the Hamiltonian (\ref{IGTHamil}) and the action (\ref{eq:z2action}). However, as
their physical import is very similar it is useful to keep this notation.

We next rewrite the partition function as:
\bea
&Z&(K,J) = (\cosh K)^{N_p} (\cosh J)^{N_l} \times \\ \nonumber
&\mathrm{Tr}_{\sigma,\tau}& \left[\prod_p(1+ \tanh K \prod_{l \in \partial p} \sigma_l )\right]
\left[\prod_l(1+ \tanh J  \sigma_l \prod_{s \in \partial l} \tau_s )\right] \ ,
\eea
where $N_{p,l}$ are the number of plaquettes and links, respectively, in the
lattice.
Multiplying out the products in the brackets and performing the trace
annihilates all terms in which any $\tau_s,\sigma_l$ appears an odd number
of times. For $J=0$, this allows only for the presence of closed
surfaces, a surface being defined as containing a given plaquette $p$ if
the factor of $\prod_{l \in \partial p} \sigma_l$ appears in the
sum.  Switching on a non-zero $J$ allows the surfaces to have free
edges. Thus we have rewritten our gauge theory as a statistical mechanics
of surfaces with edges as promised.
In this formulation each area element of the surface has a bare surface tension
$\tanh K$, and each edge a bare line tension $\tanh J$. The actual geometrical
properties of the different regimes and phases are determined by the renormalization of these
quantities due to the entropy of the surfaces and edges.

\begin{figure*}
\scalebox{0.5}{\includegraphics{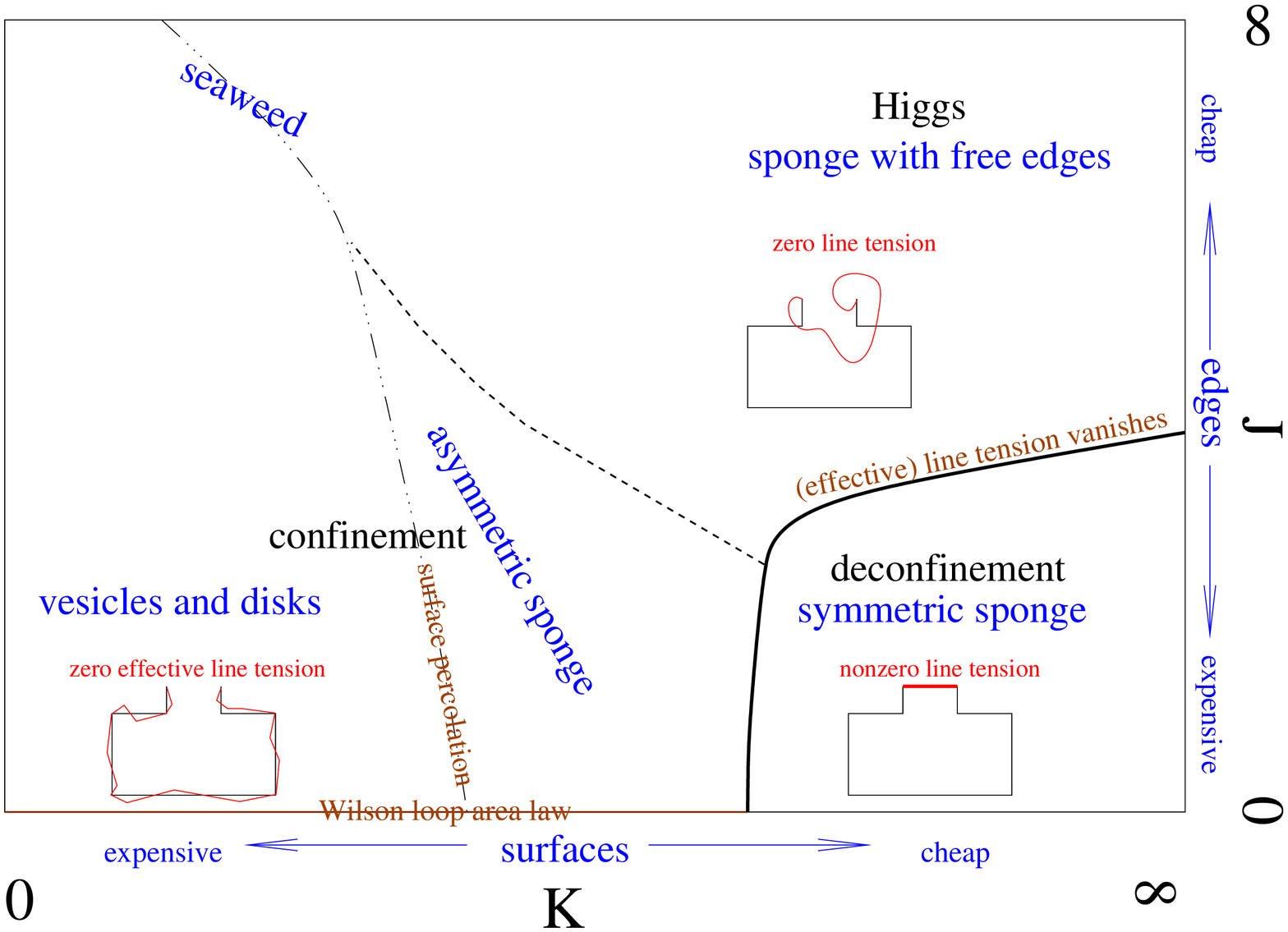}}
\caption{Schematic phase diagram of the $Z_2$ gauge theory coupled to $Z_2$ matter (\ref{eq:z2action}), interpreted in analogy to sponge phases (see text).
The inscribed cartoons capture the behavior of the Huse-Leibler horseshoe (\ref{eq:horseshoe})
and allow detection of an effective line tension.}
\label{fig:ihphase}
\end{figure*}

The phase diagram of the problem is sketched in Fig.~\ref{fig:ihphase}
and includes more structure than is significant for the gauge theory as
we explain below. We direct the reader's attention to the following features.

\noindent
{\bf Pure gauge theory:} As already noted, for $J=0$, the system excludes
edges. All surfaces are closed, which allows a consistent
definition of an inside and an outside.  For small $K$, the total
surface area remains small, and the volume of the inside of the
surfaces is tiny, whereas the outside fills most of space. In this regime
there is a macroscopic surface tension which is in one-to-one correspondence
with an area law for the Wilson loop, as the reader can check by tracking
the latter through our mapping.

At a larger $K=K_g$, the inside of the surfaces percolates as well.
This, first, transition is purely geometrical and thus does not produce
thermodynamic or local operator singularities, e.g. the surface tension is
analytic across the boundary. In the language of membranes, the
percolating phase is called the asymmetric sponge, as inside an
outside have unequal volumes.  The transition is dual to the one
encountered in the Ising models in $d=3$, where the minority phase
starts percolating at a temperature below the thermodynamic $T_c$.

At $K_c>K_g$, there is a thermodynamic transition which corresponds
to deconfinement in the pure gauge theory. This is signaled by the vanishing of
the macroscopic surface tension: In the Wilson loop acquires a perimeter law, while for the sponge the inside and outside volumes become equal
whence the new, deconfined, phase is labeled a ``symmetric sponge''.

\begin{figure}
\scalebox{0.4}{\includegraphics{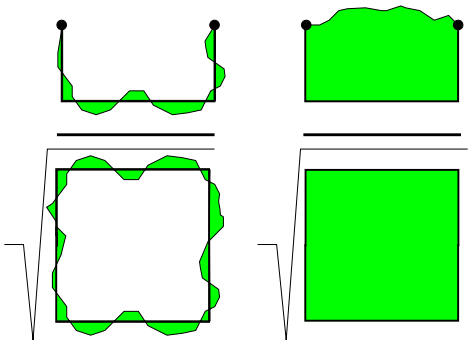}}
\caption{The line tension/Fredenhagen-Marcu diagnostic is 
the ratio of a half Wilson
loop (numerator) to the square root of a Wilson loop. Fat points 
denote matter insertions. 
On the left we give a cartoon of  the dominant behavior in the 
confined phase and on the right the dominant behavior in the deconfined phase.}
\label{fig:surfacetension}
\end{figure}

\noindent
{\bf Regime with heavy dynamical matter:\footnote{``Heavy'' refers 
to the large creation cost, $2 \Gamma_m$, rather than a small bandwidth, 
$\propto J$, of the matter particle.}} Turning on $J$ causes edges to
appear. Even at small $J$, where the bare line tension is large, edges have
the qualitatively important effect of causing the macroscopic surface tension
to vanish whence the Wilson loop exhibits a perimeter law for any $K$ and
$J>0$. However, the phase transition between the asymmetric and symmetric sponges
is not destroyed at small $J$ even if the labeling is now problematic as inside and outside
are not distinct. There continues to be an intuitive distinction in the tension
associated with surfaces between the two phases as illustrated in the
bottom panels in Fig.~\ref{fig:surfacetension}. In the symmetric sponge
phase, the Wilson loop on the right follows a perimeter law despite
the large area present, whereas on the left, a perimeter law arises
because dynamical matter allows the surface area to be proportional to
the perimeter of the loop by introducing an edge that runs along it.

\noindent
{\bf Regime with light dynamical matter:} As the bare line tension,
$\tanh J$, is increased further, edges proliferate. Of most interest
to us, this has the effect of terminating the symmetric sponge phase
and driving the system into the ``sponge with free edges'' which is
the Higgs region of the gauge theory in the present formulation. This
transition is most vivid at $K=\infty$, where surfaces drop out and we
are left with the statistical mechanics of loops alone. In fact these
loops the high-temperature expansion of the Ising model, with a vanishing line
tension at $J_c$ signalling the onset of ferromagnetic order, a
thermodynamic transition which separates the deconfined from the
Higgs phase.\footnote{Note that there is no additional geometrical
transition for the edges at a $J_g>J_c$ even in the self-dual case
$d=3$; rather, it will show up as a transition in the appropriate dual
variables.}

The sponge with free edges is accessible, without crossing a thermodynamic
phase transition, from both the asymmetric sponge and the vesicles and disks
regions, in keeping with the continuity between the Higgs and confinement
regions.
However, the appearance of the
surfaces nonetheless varies drastically. Deep in the Higgs phase, long
edges frame large surfaces. By contrast, in the `seaweed'
regime, edges are still cheap and the tradeoff is between edge entropy
and the cost of the surfaces attached to the edges, leading to lines
joined by (lattice dependent) surfaces of close to minimal area.

\section{The Huse-Leibler horseshoe}

Having described the phase diagram of the $Z_2$ gauge theory in terms of our
system of surfaces with edges, let us now turn to the task of distinguishing
the symmetric sponge/deconfined phase from all others. Intuitively, we need
a quantity which teases out whether the perimeter law of the Wilson loop is
fundamentally due to a vanishing surface tension, or instead due the absence of
a surface spanning it {\it and} whether there is an underlying large line
tension. Somewhat non-intuitively this can be accomplished by simply defining
an appropriate {\it line} tension.

To this end consider the geometrical object inscribed in
Fig.~\ref{fig:ihphase}. It is a Wilson loop broken open, so that its
total length remains fixed but so that gap of size $g$ opens up. To
keep the quantity gauge invariant, it is necessary to terminate the
open strings by a matter term, so that the Huse-Leibler horseshoe \cite{HuseLeibler}
is defined by the quantity
\bea
\label{eq:horseshoe}
H (L,g) =  \left<  \tau_s \left( \prod_{l \in  C_H(s,s')} \sigma_l \right) \tau_{s'}  \right>,
\eea
where the product is taken over the links $C_H(s,s')$ constituting the 
horseshoe between $s$ and $s'$.

The basic point is that, if surfaces are cheap and edges are expensive---the situation
prevailing in the symmetric sponge---the system will cover most of the
horseshoe with a surface and then close the gap with an edge at a cost in free
energy of $\exp(-\lambda g)$ due to the tension of the line closing the gap (red
line in Fig.~\ref{fig:ihphase}). By contrast, if surfaces are expensive or/and
edges are cheap which is the situation elsewhere in the phase diagram, no such
dependence on $g$ arises.
As a practical matter, to separate out the dependence on $g$ from that on the linear size of
the horseshoe, $L$, we need to study this quantity in the limit $L \gg g \gg 1$
and then check for an effective line tension $\lambda$ in the expectation value of
the horseshoe.

An elegant alternative is to remove the $L$-dependence by dividing the
horseshoe by the Wilson loop of same size $L$, as the leading $L$
dependence of the two is set by the thermodynamic properties of
surface and edges. Once this has been done, the resulting ratio will
either vanish as $\exp(-\lambda g)$ in the symmetric sponge, or
approach a constant.
Indeed, one does not even need to introduce a separate length scale $g$
altogether. One can instead decide to cut the Wilson loop in two. For
a square loop of side length $L$, as shown in Fig.~\ref{fig:surfacetension},
this yields two rectangular
loops of size $L/2 \times L$, one long side of which is open and
now takes over the role of the gap; again, a $\tau$ needs to be
attached to each open end. The result is the line tension diagnostic ratio
\begin{equation} \label{ratiodefinition}
R(L) = {W_{1/2}(L) \over \sqrt{W(L)}}=
{\langle \tau_s (\prod_{l \in C_{1/2}} \sigma_l) \tau_{s'} \rangle
\over \sqrt{\langle \prod_{l \in C} \sigma_l \rangle}}
\end{equation}
where the curves $C_{1/2}$ and $C$ define the half-Wilson loop $W_{1/2}$ and full Wilson loop $W$ respectively.
$R(L)$ has the desired asymptotic behavior:
\bea
&\lim_{L \rightarrow \infty} R(L)& =  0 \ \ {\rm deconfined \ phase} \nonumber \\
&\lim_{L \rightarrow \infty} R(L)&  \neq  0  \ \ {\rm otherwise} \ .
\label{eq:Rinfty}
\eea

Before we work further with $R$, three comments are in order. First, along the $K=\infty$
axis this ratio is simply the spin-spin correlation function of the Ising model evaluated
between the endpoints of the half Wilson loop and it distinguishes the two phases as asserted.
Second, this diagnostic fails along exactly one line---when $J=0$, but fortunately the
Wilson loop itself is available exactly in that case. Third, we have checked by Monte Carlo
and in the small $J$ expansion, that $R$ has the claimed behavior.

\section{Operator Formulations and the Fredenhagen-Marcu Order Parameter}

The line tension diagnostic introduced in the last section can be oriented 
in an arbitrary direction for the Lorentz invariant example studied thus far. To 
obtain an operator formulation, we must pick a particular orientation. Here we 
consider three and gain some insight into the operation of the diagnostic.

\subsection{Equal time formulation}
\label{ss:EqTimeForm}

The simplest option is to orient the variables in the ratio so that they all lie in the same time slice of the path integral as in Fig.~\ref{fig:equaltime}. 
Then the transcription to the Hamiltonian formulation is straightforward and we obtain from the definition in Eq.~(\ref{ratiodefinition}):
\begin{equation} \label{ratiodefinition-op}
R(L) \equiv {W_{1/2}(L) \over \sqrt{W(L)}} =
{\langle G | \tau^z_s (\prod_{l \in C_{1/2}} \sigma^z_l) \tau^z_{s'} | G \rangle
\over \sqrt{\langle G | \prod_{l \in C} \sigma^z_l | G \rangle}} \ ,
\end{equation}
where $|G \rangle$ is the ground state.
As written, at $T=0$, $R(L)$ clearly measures a property of the ground state wavefunction but it is straightforward to use it at $T >0$ with thermal averages replacing ground state ones. Transcribing
our previous analysis, $R(\infty)$ vanishes if the equilibrium gauge field fluctuations are weak and the matter is uncondensed but not otherwise. This turns out to be the form of the diagnostic
which can be used most generally for reasons that we explain later.

\subsection{Fredenhagen-Marcu Order Parameter}
\label{ss:FMOP}

\begin{figure}[ht!]
    \label{fig:operatorforms}
    \begin{center}
        \subfigure[\ Equal time diagnostic]{
            \label{fig:equaltime}
            \includegraphics[width=0.4\textwidth]{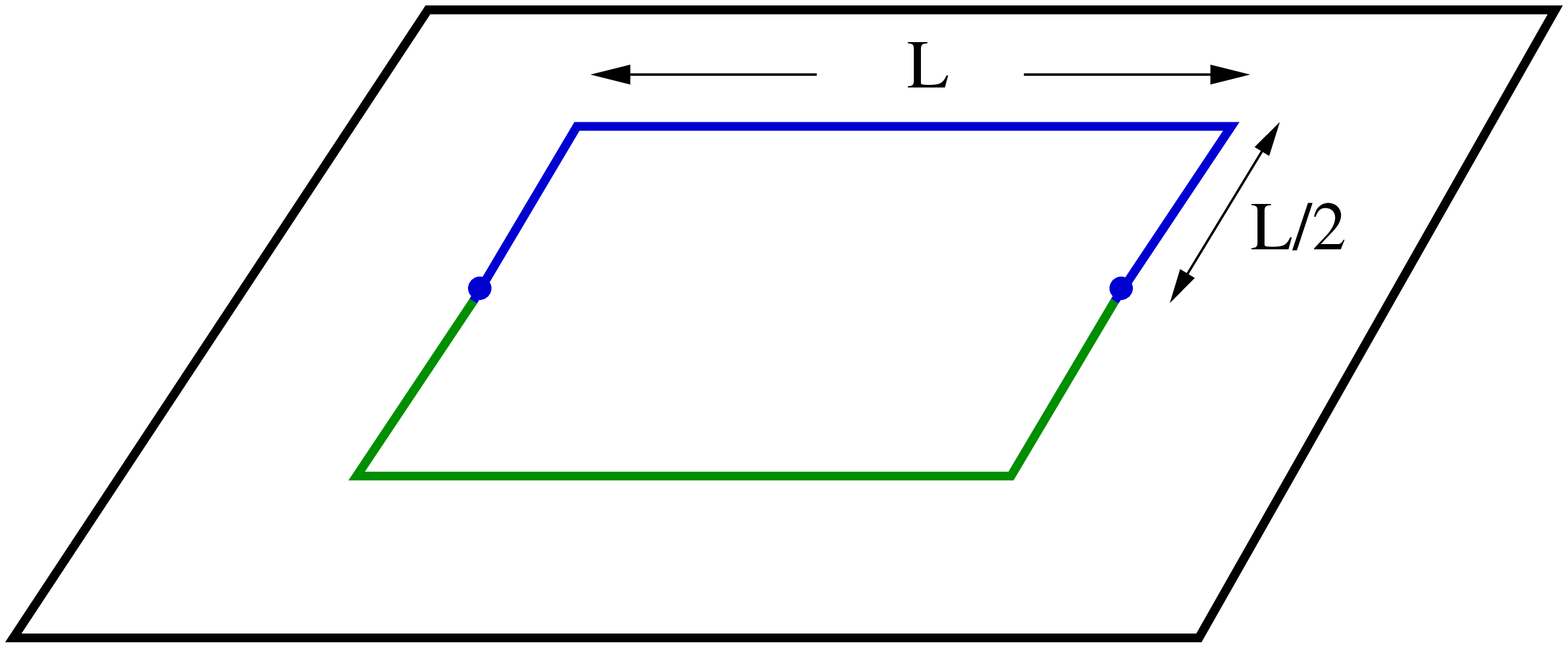}
        } \\
        \subfigure[\ Fredenhagen-Marcu ``Order Parameter'']{
           \label{fig:FM}
           \includegraphics[width=0.4\textwidth]{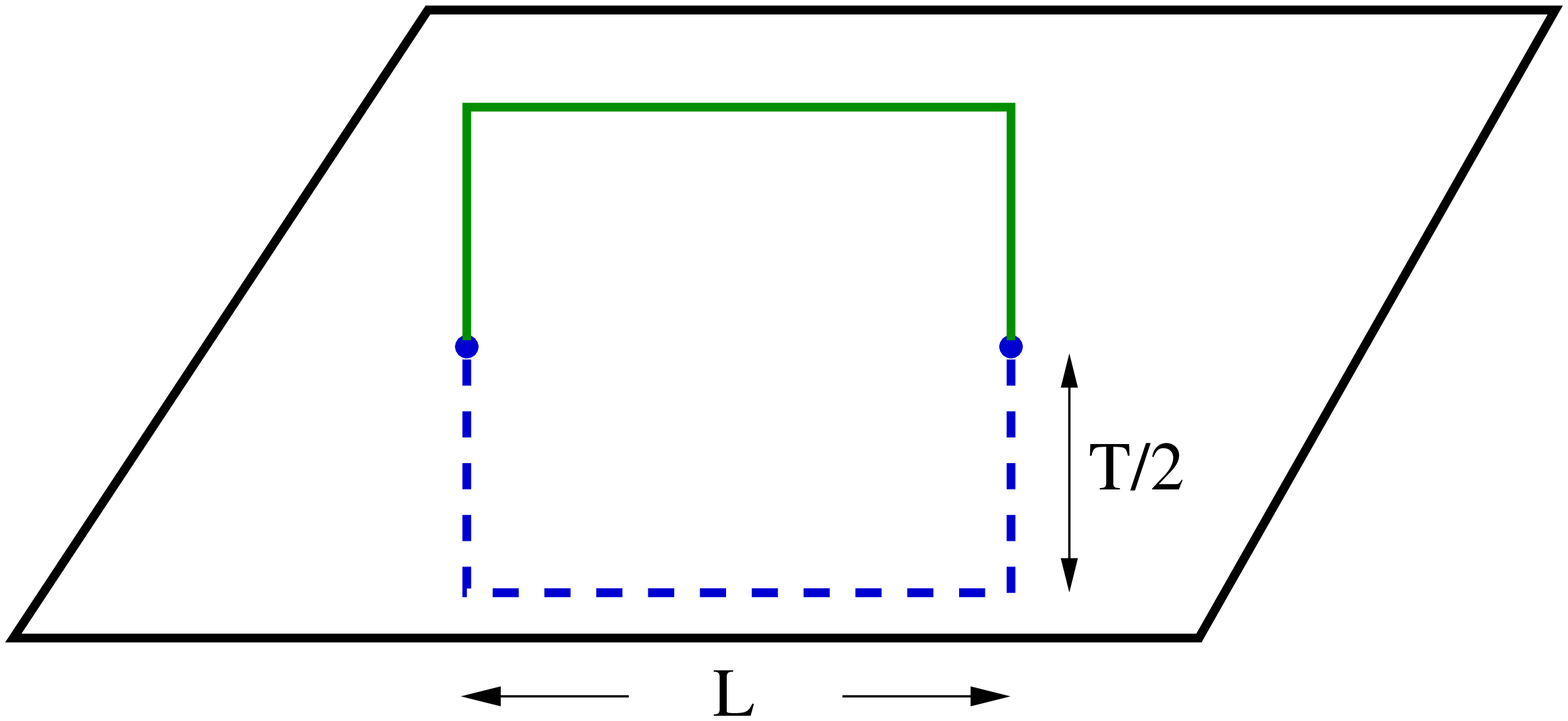}
        } \\ 
        \subfigure[\ Spinon delocalization diagnostic]{%
            \label{fig:spinondeloc}
            \includegraphics[width=0.4\textwidth]{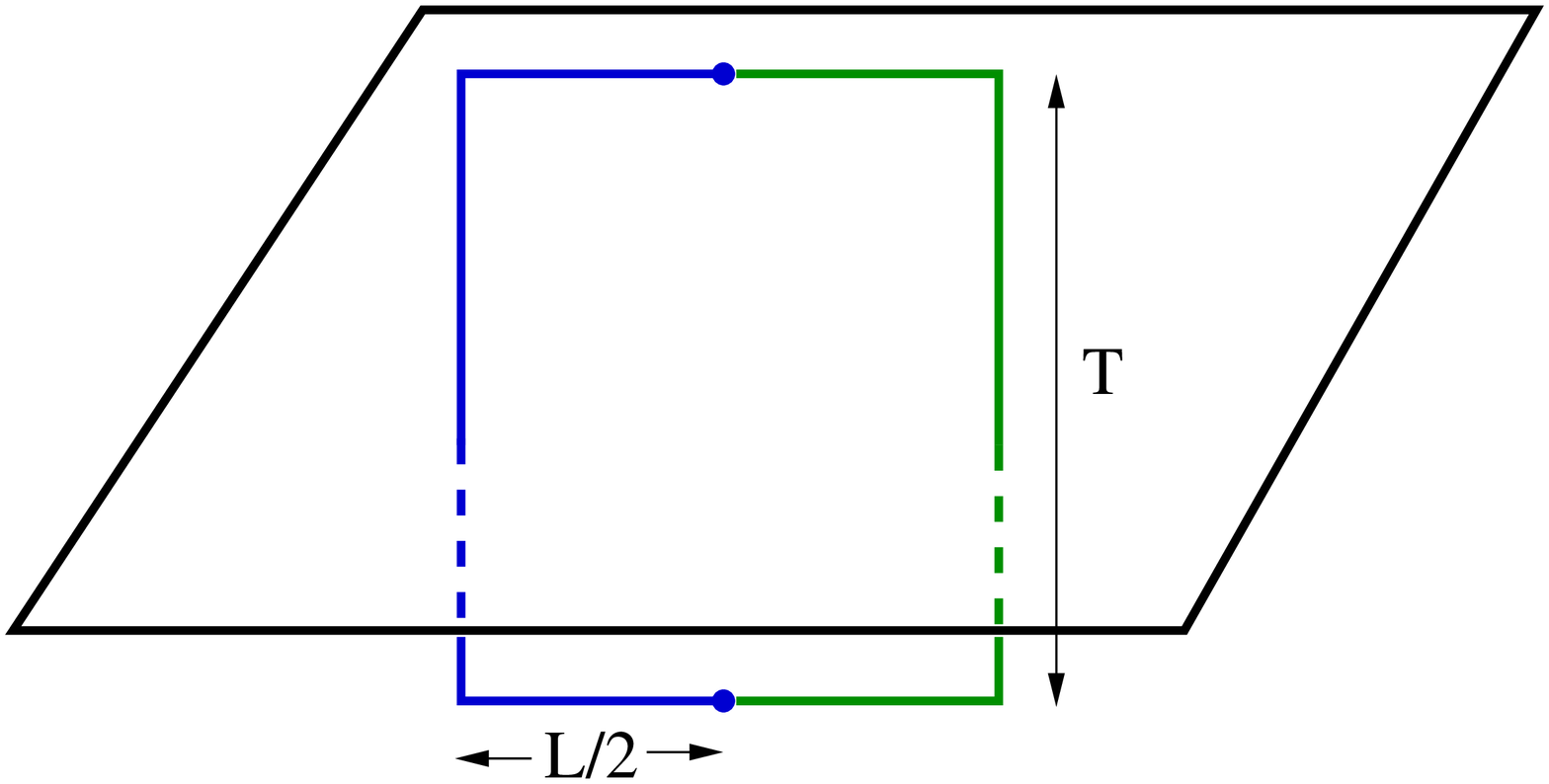}
        }%

    \end{center}
    \caption{%
         Different operator formulations of the deconfinement diagnostic (\ref{ratiodefinition}). The blue curve and
        terminating points define the locus of numerator and the green curve marks
        the needed completion for the denominator. 
             }%
\end{figure}

Now let us orient the operators so that they lie in a plane containing the imaginary time axis with the endpoints of the half Wilson loop at the same time (see Fig.~\ref{fig:FM}) which we take to be zero for convenience. Then we can write the operator expression
\beq
W_{1/2} = \langle G | \tau^z_s \tau^z_{s'} \prod_{l \in C_{s s'}} \sigma^z_l (-T / 2) | G \rangle
\eeq
where $T=c L$ up to multiplication by a velocity $c$, here set to 1, $C_{s,s'}$ is the linear path between $s$ and
$s'$, and
\beq
\sigma^z_l (-{T / 2}) = e^{-HT/2} \sigma^z_l e^{-HT/2}
\eeq
and $e^{-H}$ is the transfer matrix for one step in the time direction---with the implied meaning in the time continuum limit.
[In the process of passing to the Hamiltonian setting we have, as usual, set $\sigma^z=1$ along the temporal links.]

Similarly, we can write the denominator as
\beq  \label{eq:FMdenominator}
W = \langle G |  \prod_{l \in C_{s s'}} \sigma^z_l (+T / 2)
\prod_{l \in C_{s s'}} \sigma^z_l (-T / 2)  | G \rangle \ .
\eeq

Now define,
\beq
|s s' \rangle = \tau^z_s \tau^z_{s'} \prod_{l \in C_{s s'}} \sigma^z_l (-T / 2) | G \rangle
\eeq
as the candidate two spinon state constructed from the ground state. Thus we see that
\beq
\label{eq:FMop}
R(L) = { \langle G | s s' \rangle  \over \sqrt{  \langle s s' | s s' \rangle} }
\eeq
i.e. it is the overlap between the ground state and the normalized two spinon state. In this form, our diagnostic was first discovered by Fredenhagen and Marcu \cite{FMop1,FMop2} and they proved that it does show the asymptotic behavior (\ref{eq:Rinfty}) in the $Z_2$ gauge theory with matter. In subsequent work they also considered the equal time form  (\ref{ratiodefinition-op}).

It follows from our previous considerations that
\begin{itemize}
\item in the deconfined phase the candidate two spinon state is orthogonal to the ground state in the limit $c L=T \rightarrow \infty$, and has finite energy and thus there are free spinons in the spectrum
\item in the confined-Higgs phase the candidate two spinon state is {\it not} orthogonal to the ground state and thus there are {\it no} free spinons in the spectrum
\end{itemize}
In this fashion we see that in this interpretation, $R(\infty)$ distinguishes between the phases based on the existence of free spinons in the spectrum.

Finally, it is instructive to see how these limits work directly in the Hamiltonian formalism. Consider the sequence

\renewcommand{\labelenumi}{(\alph{enumi})}
\begin{enumerate}
\item the gauge-noninvariant string $\prod_{l \in C_{s s'}} \sigma^z_l(-T/2)$ acts on the ground state leading to a state with a modified constraint at $s$ and $s'$
\item the system evolves/relaxes for an imaginary time $T/2$
\item the gauge-noninvariant product $\tau^z_s \tau^z_{s'}$ acts on the state at the end of the last step and leads to a final gauge invariant state
\end{enumerate}

Deep in the deconfined phase (a) and (b) leave an additional electric 
flux string that meanders between the sites $s$ and $s'$ and (c) adds the 
matter needed to create a proper two spinon state.
Conversely, deep in the confined phase, (b) eliminates the string introduced by (a) in favor of fluxless charges at $s$ and $s'$ which are then eliminated in (c) leaving the ground state behind.

\subsection{Spinon delocalization}
\label{ss:SpDel}

The two  interpretations provided above are sufficient for our purposes in the rest of the 
paper, but some further insight can be gained by considering a third one: Now 
we orient the half Wilson loop so that $s$ and $s'$ are separated 
by imaginary time $T$ as shown in Fig.~\ref{fig:spinondeloc}.

With this choice, the numerator becomes  the expectation value,
\begin{equation}
\label{eq:FMdenominator}
W_{1/2} = \langle G | \tau^z_s(T/2) \prod_{l \in C_{0s'}} \sigma^z_l (T / 2)
\tau^z_{s}(-T/2) \prod_{l \in C_{0s}} \sigma^z_l (-T / 2) | G \rangle
\end{equation}
while the denominator is given by the same expression exhibited previously.

To understand why the ratio $W_{1/2}/\sqrt{W}$ behaves differently in the two phases consider the intermediate states produced by imaginary time evolution. In the deconfined phase
\begin{itemize}
  \item $W_{1/2}$ couples to a state with one defect site at $s'$ and a delocalized spinon, while
  \item $W$ couples to a state with two widely separated defect sites.
\end{itemize}
Hence the ratio is controlled by intermediate states with different energies and scales as
\beq
\frac{ e^{-(E_{\rm defect} + E_{\rm spinon}) T}}{\sqrt{ e^{-(E_{\rm defect} + E_{\rm defect}) T }}}
\eeq
and thus vanishes at large $cL=T$.

By contrast, in the confined phase, both operators couple to eigenstates with localized spinons, at $s'$ for $W_{1/2}$ and at both $s,s'$ for $W$, whence the ratio tends to a constant. It follows then, that in this version the diagnostic is sensitive to the existence of a delocalized spinon state in the deconfined phase.

\section{Fluctuating constraints and choice of operator formulation}
\label{sec:flucconstr}

In a condensed matter context, a constraint such as the one encoded 
in Eq.~\ref{eq:IGTconstraint} is not microscopic (or ``fundamental'') but rather of emergent origin, that is to say it is a legacy of higher energy terms in the Hamiltonian. The constraint is thus not inviolable---rather, states outside the ``gauge invariant'' or ``physical'' sector are associated with a large but finite
energy penalty set by a scale $U$. To put it differently, in the context of gauge theory it is
assumed that ``unphysical'' states are really unphysical---they are forced on us by the challenge
of handling the constraint---but in condensed matter setting they are certainly physical but are
instead high energy states.

This distinction has an important consequence for the operator versions of our diagnostic, e.g.
the Fredenhagen-Marcu version (\ref{eq:FMop}). As noted in the discussion following its
introduction, it contains a two spinon state creation operator that can be factored into two
pieces,
$$
\tau^z_s \tau^z_{s'} \ \ {\rm and} \ \ \prod_{l \in C_{s s'}} \sigma^z_l (-T / 2)
$$
which are not individually gauge invariant, although their combination is. Now the second
piece is sensitive not only to matrix elements of $H$ between gauge invariant states but {\it also}
to matrix elements of $H$ involving gauge non-invariant states. Indeed our verbal explanation of the
creation of the two spinon state referred specifically to imaginary time evolution in the
gauge non-invariant sector where the constraint was violated at two sites. For the specific choice of Hamiltonian (\ref{IGTHamil}) all is well---the detour into the 
gauge non-invariant sector introduces
no physics inconsistent with the proper operation of the diagnostic.  Of course this had
to be the case as we got this form working from a manifestly Lorentz invariant Euclidean
path integral. We now show that this does not {\it have} to be the case by explicitly exhibiting
an example with a soft constraint where the Fredenhagen-Marcu version of $R(L)$ fails to
diagnose deconfinement. The example is somewhat contrived but it is intended only to
function as a cautionary counter-example.

Consider modifying our starting $Z_2$ gauge theory Hamiltonian (\ref{IGTHamil}) to $H_f$:
\begin{equation}
\label{eq:IGTHamFluc}
H_f = H + {U \over 2} \sum_s (G_s-1)^2 - \lambda \sum_s \left( G_s - 1 \right) \tau^x
\end{equation}
and set $J$, $\Gamma \to 0$ for now. Now we do {\it not} implement the constraint (\ref{eq:IGTconstraint}) on the Hilbert space, instead a suitably large value of $U$ does
this in the low energy sector. With that ordering of energy scales, the ground state at
small $\lambda \ll U$
\begin{equation}
\mid G \rangle = \left( \prod_s \frac{1}{2} \left( G_s - 1 \right) \right) \mid \left\{ \sigma^z = 1 \right\} \rangle \otimes \mid \left\{ \tau^x = 1 \right\} \rangle
\label{eq:GGs}
\end{equation}
has precisely the same form as that of the toric code. However, there is now
a change of meaning, as microscopically both the link and site variables are physical while
in the toric code the site variables can be gauged away. Yet, at low energies in our model
the same reduction takes place and so it has as many degrees of freedom and upon our choices
of parameters exhibits the same, but now emergent, $Z_2$ order.

Now consider the excitations of our model. The gauge invariant excitations preserve $G_s=1$ 
at
all sites and are visons with an energy $2K$ and spinons with an energy $2 \Gamma_M$. Note
that their energies are independent of $\lambda$; indeed, all gauge invariant states have (by
construction) energies independent of $\lambda$. The gauge non-invariant excitations 
involve violating
the constraint by setting $G_s=-1$, and also come in two species: spinonic defects with
energy $U+2 \Gamma_M - 2 \lambda$ obtained by setting $\tau^x=-1$ at a site and gauge defects
with energy $U + 2 \lambda $ obtained by acting with a string $\prod \sigma^z_l$ extending
from a given site to infinity. Observe that the energies of the two defect states cross when
$U \gg \lambda > \Gamma_M/2$ even though the spectrum of the entire set of lower energy
gauge invariant states is unchanged. [For the Hamiltonian (\ref{IGTHamil}) the spinonic and 
gauge
defects have energy $2 \Gamma_M$ and $0$ respectively. While there is no explicit energy
penalty for gauge non-invariant states, the separate requirement that the ground state obey
(\ref{eq:IGTconstraint}) eliminates the degeneracy that would be present otherwise.]

In our previous discussion we argued that the action of
$\prod_{l \in C_{s s'}} \sigma^z_l (-T / 2)$ was to create a string stretching between sites
$s$ and $s'$. This works as long as the state with two gauge defects is the ground state in
the sector with two defect sites. Once the defects cross and the two spinonic defects become
the ground state, this is no longer true. As a technical matter, the crossing alone is not
enough, we also need to turn $J$ on to a small value so that imaginary time evolution connects the two states. Once this is done, $R(\infty) \ne 0$ even though we are clearly in the deconfined
phase and the diagnostic fails in this form.

However the diagnostic still works in its equal time form so the lesson is that in general
condensed matter settings the formulations presented in the previous section are not fully equivalent: whereas the equal--time formulation (Sec. \ref{ss:EqTimeForm}) is robust,
the other pair of formulations (Sec. \ref{ss:FMOP},\ref{ss:SpDel}) need not be.

\section{Finite temperature topological order in $d \ge 3$}
\label{sec:fintemp}

We now return to our main development and consider the use of our diagnostic to establish
the finite temperature phase structure of the $Z_2$ gauge theory where contrasting answers
can be found even in the recent literature \cite{Gliozzi, Claudios3d,SentFish3d}.
We begin by considering the $T>0$ physics of the toric code,  $J=\Gamma=0$ in Eq.~\ref{IGTHamil}.
It can be shown (see Appendix~\ref{app:kita}) that at this point in $d+1$ space-time dimensions
the quantum partition function at all temperatures is proportional, up to a simple multiplicative
analytic factor, to that of the pure $Z_2$ gauge theory {\em without matter} defined on a $d$
dimensional Euclidean lattice.

From what is known about the latter theory, one can immediately read off that topological order will persist to finite temperature for all $d\geq3$, but not below. Further we see that, despite the
presence of matter, the transition between the low temperature deconfined phase the the high
temperature phase can be diagnosed by the Wilson loop which goes from a perimeter law to an
area law between the two phases. 

This physics has to do with the gauge excitations.
For $d=2$, finite energy point-like visons appear at an exponentially small but non-zero density
at any $T>0$ destroying the deconfined phase. As the Wilson loop measures the averaged
parity of the enclosed vorticity, it acquires an area law as a result. 

By contrast, this mechanism is not operative in higher dimension because the gauge/magnetic excitations  now appear in a higher-dimensional version---loops in $d=3$---whose characteristic size, $\xi_m$, vanishes as $T\rightarrow0$. As the Wilson loop now counts the parity of flux loops which link it, it is sensitive only to those within $\xi_m$ of its boundary, whence the perimeter law and the deconfined phase survive at sufficiently small temperatures. 

At higher temperatures there is a phase
transition, e.g in $d=3$ where the vortex loops unbind.
The basic physics of vortex loops discussed above indicates that the finite temperature
deconfined phase should be stable beyond the Kitaev point in $d \ge 3$ as argued in
\cite{SentFish3d, Claudios3d}.  However, once matter becomes dynamical the Wilson loop exhibits a
perimeter law at any $T$ and other standard diagnostics fail as well.

Fortunately, our diagnostic continues to work at $T>0$. At small temperatures one can work
perturbatively at small $J$, when the dynamical matter is heavy. As at $T=0$, the basic
result is that in this limit the behavior of $R(L)$ follows from that of $W(L)$ in the
absence of dynamical matter. As the latter exhibits a transition between perimeter and
area dependence, $R(\infty)$ transitions between vanishing and developing an expectation
value of $O(L^0)$. In somewhat more detail, one can examine the Euclidean path integral, now
finite in the (large) imaginary time direction at low temperatures much as in Section \ref{sec:spongemapping}. As before, a large line-tension between any pair of charges survives,
and it continues to be preferable to minimize free edges at the expense of adding the (cheap) surface, leading to a vanishing of $R(\infty)$.

At high temperatures, $T\rightarrow\infty$, it is more convenient to work directly in the
Hamiltonian formulation and the high-temperature expansion. In this expansion, one obtains polynomials in powers of $\beta=1/T$ upon expanding $\exp(-\beta H)$. Non-vanishing terms in the expectation values appearing in Eq.~\ref{ratiodefinition} are only obtained if each unpaired $\sigma^z, \tau^z$ is matched by the polynomials. The cheapest (in powers of $\beta$) way of doing this is to retrace the perimeter of $W$ and $W_{1/2}$. This leads to the same leading power of $\beta$ in the
expectation value of each, which thence cancel in the expression for  $R(\infty)$. Consequently
$R(\infty)$ is non-zero at sufficiently high temperatures and we that there is a phase transition
between the high and low temperature phases.

\section{Beyond $Z_2$}
\label{sec:beyond}

\subsection{Gauge fields with fundamental matter}

Thus far our discussion has focused on the simplest case, that of standard $Z_2$ gauge theory with
$Z_2$ matter where there are two phases to distinguish---the deconfined phase and the confined-Higgs phase. Our deconfinement diagnostic generalizes straightforwardly to lattice gauge theories based
on other gauge groups where there are similarly only two phases to distinguish, to $Z_n$, $U(1)$
and to Yang-Mills theories based on non-abelian groups. Indeed, the last named case was the primary spur to Fredenhagen and Marcu's work. In all of these cases $R$ has the same structure with
appropriate replacement of the matter and gauge fields.

For example, a compact $U(1)$ gauge theory with charge $q=1$ matter is defined by a space time action analogous to (\ref{eq:z2action})
\bea
\label{eq:u1action}
-S=K \sum_p \prod_{l \in \partial p} U_{\tilde l(l)}  + J \sum_{l; s,s'\in \partial l} \tau_s U_l \tau_{s'} + {\rm h.c.}
\ ,
\eea
where the $U_l=\exp(i A_l)=U^\ast_{-l}$ are now $U(1)$ valued gauge fields living on the links
defined
in the standard orientation from one sublattice to the other, and the
$\tau_s=\exp(i \phi_s)$ are the corresponding matter fields residing on the sites. Unlike the
$Z_2$ theory, we now need to pay attention to the orientation of various link variables. The links
in the second term are traversed in the standard orientation while in the first term the loops
are taken anticlockwise with $\tilde l (l) = \pm l$ if the link is traversed in, or opposite to, its standard
orientation respectively.

\begin{figure}[h]
\epsfxsize=2.3 in \centerline{\epsfbox{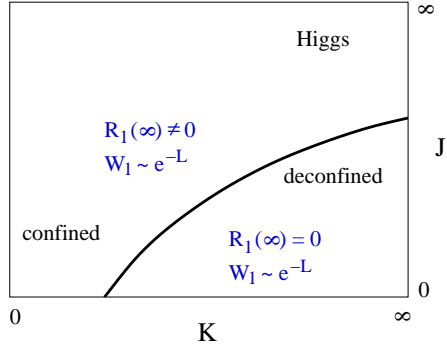}}
\caption{Sketch of the phase diagram of the $U(1)$ gauge theory with
charge one matter fields. The behaviour of the line tension diagnostic $R_1$ and Wilson loop $W_1$ is indicated for each phase.} 
\label{fig_U1q1_phase_diag}
\end{figure}

At zero temperature in spatial dimension $d\geq 3$ this theory exhibits both a confined and a
deconfined phase,\cite{FradkinShenker} see Fig.~\ref{fig_U1q1_phase_diag}.
Absent matter, $J=0$, the oriented Wilson loop
\bea
W_1[C] =  \langle \prod_{l \in C}  U_{\tilde{l}(l)} \rangle
\label{eq:u1q1wloop}
\eea
is capable of distinguishing
between these two phases, exhibiting an area law in the former, and a perimeter law in the latter,
with logarithmic corrections signaling the Coulomb interaction \cite{kogutRMP}.

This phase structure persists when matter with charge $q=1$ is turned on. However,
as in the case of the $Z_2$ theory above, the
Wilson loop fails as a diagnostic in this setting. By  analogy to
Eq.~\ref{ratiodefinition}, we construct the appropriately generalized diagnostic
ratio,
\bea
R_1(L) = {W_{1/2}(L) \over \sqrt{W(L)}}=
{\langle \tau^\dagger_s (\prod_{l \in C_{1/2}} U_{\tilde{l}(l)} \tau_{s'} \rangle
\over \sqrt{\langle \prod_{l \in C} U_{\tilde{l}(l)} \rangle}} \ .
\label{eq:u1q1diag}
\eea
As before the phases are distinguished by the asymptotic behavior of $R_1(L)$,
\bea
&\lim_{L \rightarrow \infty} R_1(L)& =  0 \ \ {\rm deconfined \ phase} \nonumber \\
&\lim_{L \rightarrow \infty} R_1(L)&  \neq  0  \ \ {\rm otherwise} \ .
\label{eq:R1infty}
\eea

We now turn to the application of the diagnostic to cases with more than one phase and to problems
phrased in more standard condensed matter settings.

\subsection{U(1) gauge theory with charge $q$ matter}

To obtain a richer phase structure we consider a $U(1)$ gauge theory coupled to matter fields
that carry one or more higher (non-fundamental) charges $q\geq 2$. In this setting it is useful
to consider the set of charge-$q$ operators
\bea
W_q(L) =  \langle \prod_{l \in C} U^q_{\tilde{l}(l)} \rangle ; \ \ \
R_q(L) = {\langle \tau^\dagger_s (\prod_{l \in C_{1/2}} U^q_{\tilde{l}(l)}) \tau_{s'} \rangle
\over \sqrt{\langle \prod_{l \in C} U^q_{\tilde{l}(l)}) \rangle}} \ .
\label{eq:u1qdiag}
\eea
whose structure indicates that $q$ electric flux quanta emanate from
a given charge.

\begin{figure}[h]
\epsfxsize=2.3 in \centerline{\epsfbox{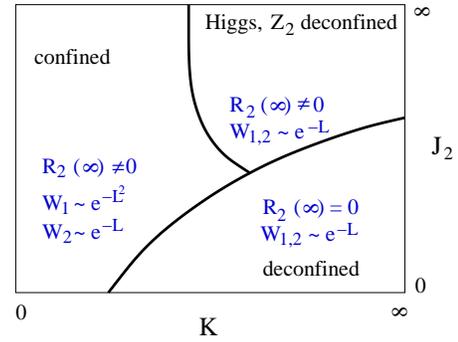}}
\caption{Sketch of the phase diagram of the $U(1)$ gauge theory with
charge two matter fields. Charge-$q$ line tension diagnostic $R_q$ and Wilson loop $W_q$ are defined in Eq.~\ref{eq:u1qdiag}.}
\label{fig_U1q2_phase_diag}
\end{figure}

As our first example we consider the well known phase diagram of a single matter field of
charge $q=2$. This contains, as shown in Fig.~\ref{fig_U1q2_phase_diag}, in addition to the deconfined phase of the $U(1)$ field, a Higgs phase which corresponds to the deconfined phase of the appropriate $Z_2$ gauge theory and a distinct completely confined phase
\cite{FradkinShenker}.
Due to the lack of $q=1$ matter, we can make some progress using Wilson loops alone. 

Of the three phases, two ($U(1)$ and $Z_2$ deconfined) exhibit a perimeter law in $W_1$ while
one ($U(1)$ confined) exhibits an area law. The key point is that the $q=2$ dynamical matter
cannot break up the confining string that stretches between two test $q=1$ sources in the
confined phase. By contrast, $W_2$ exhibits a perimeter law everywhere and distinguishing the
two deconfined phases now requires that we evaluate the diagnostic $R_2$. As this tests for
the presence of doubly charged states in the spectrum, it vanishes {\it only} in the fully
deconfined phase.

In terms of the nature of gauge-field fluctuations, 
these distinct behaviours can be seen most intuitively by gauge-fixing $\phi_s \equiv 0$. 
Then, deep in the deconfined phase, an appropriate projection of the state $\otimes_l | A_l = 0 \rangle$ is a ground state.
In contrast, the gauge field fluctuates strongly in the confined phase, and the $A_l$ range over the full interval $[0,2\pi)$.In the
Higgs fluctuations are restricted: here, $A_l = 0$ or $A_l = \pi$ satisfy the second term of (\ref{eq:u1action}). By construction, $R_2$ is sensitive to the former, but not the latter, fluctuations, and it can thus diagnose the Higgs-confinement transition. This state of affairs is summarized in Table \ref{tab1}.

\begin{table}[h]
\begin{tabular}{lll}
\toprule
Phase & $W_1(L)$ & $R_2(L)$ \\
\colrule
Deconfined & Perimeter & 0\\
Higgs & Perimeter & $L^0$ \\
Confined & Area & $L^0$ \\
\botrule
\end{tabular}
\caption{Diagnostics for $U(1)$ gauge theory with only $q=2$ matter}
\label{tab1}
\end{table}

\begin{figure}[h]
\epsfxsize=2.3 in \centerline{\epsfbox{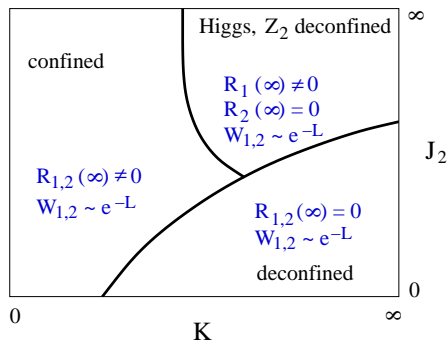}}
\caption{Sketch of the phase diagram of the $U(1)$ gauge theory with charge two as well as heavy charge one matter fields.}
\label{fig_U1q2_phase_diag2}
\end{figure}

As a second example we consider the more generic and complex situation where charges with both
$q=1$ and $q=2$ are present simultaneously. The number of parameters in the Hamiltonian makes
for a large phase diagram, and we restrict ourselves to the instructive case where the $q=1$ matter
is very heavy, {\it i.e.}\ very sparse. This then has the same physics of our last example,
and thus the same phase diagram as Fig.~\ref{fig_U1q2_phase_diag}. The new feature is that
now $W_1$ obeys a perimeter law in all three phases so that we may no longer use it to test
for the deconfinement of charge-1 objects. The solution to this is to evaluate $R_1$ instead.
The diagnostic behavior of the combination of $R_1$ and $R_2$ is summarized in Table \ref{tab2}.

\begin{table}[h]
\begin{tabular}{lll}
\toprule
Phase & $R_1(L)$ & $R_2(L)$ \\
\colrule
Deconfined & 0 & 0\\
Higgs & $L^0$ & 0 \\
Confined & $L^0$ & $L^0$ \\
\botrule
\end{tabular}
\caption{Diagnostics for $U(1)$ gauge theory with $q=1$ and $q=2$ matter}
\label{tab2}
\end{table}

We note that this combination manages to distinguish all three phases, whereas the $W_{1,2}$
fail to do so.

These deconfined phases persist to finite temperatures in dimension $d\geq 4$. The known
behavior in lower dimensions is more complicated. The $U(1)$ deconfined phase is absent
in $d=3$ at $T >0$ and in $d=2$ even at $T=0$ while the $Z_2$ deconfined phase is present
in both these cases. This complexity is correctly captured by our doublet of diagnostic ratios.

\subsection{Gauge theories of quantum magnets}

The formalism developed above can be directly applied to two classes of Hamiltonians which
arise naturally in discussions of quantum magnetism. The first are the ``slave particle''
or Schwinger boson/fermion Hamiltonians \cite{slaveCole,slaveBZA,auerarov}.
Here the spin Hamiltonian of a quantum magnet
is reformulated as a matter gauge system with spinons coupled to a gauge field. Here our
construction of $R(L)$ carries over pretty much directly. 

The second class are quantum dimer
models \cite{QDMreview, RokKiv}, now with a Hilbert space expanded to include gapped, spin-carrying monomers to represent spinons \cite{BBBSS}. In this setting, the
microscopic gauge structure arises from the hard core dimer constraint and the Wilson loop is defined as a dimer rearrangement on a loop that respects the hard core constraint \cite{nR3}. Accordingly we can again define our diagnostic.

\section{Conclusion}
We have substantially met the challenge posed at the start of this paper, that of
finding a diagnostic for deconfinement that has the form of a ground state expectation value,
i.e. which is a property of a single ground state of the system. It is also nice to find that our
reasoning, which comes from thinking about line and surface tensions in the statistical mechanics
of surfaces and edges, leads to an object discovered by lattice gauge theorists thinking about
creating states with (non) deconfined quarks. 

The next steps are to apply this formulation
to problems other than those addressed or reviewed in this paper and two come to mind. The
first is using it to detect spin liquids in systems with Heisenberg symmetry where the gauge
field involves valence bonds and there is a set of orthogonality issues to contend with. The
second is generalize the construction to non-abelian topological phases captured by the
models constructed by Levin and Wen \cite{levinwen}.

\section{Acknowledgements}

This work was supported in part by NSF grant No.~DMR-1006608 (SLS).

\appendix

\section{Kitaev's toric code}
\label{app:kita}

The model describing Kitaev's toric code
is obtained from  the Ising gauge Hamiltonian, Eq.~\ref{IGTHamil}, straightforwardly.
First, set $\Gamma=J=0$. Secondly, replace the
matter term proportional to $\Gamma_M$ by utilizing the constraint, Eq.~\ref{eq:IGTconstraint},
\beq
G_s\equiv1 \Longrightarrow \tau_s^x \equiv \prod_{l:s\in \partial l}\sigma_l^x
\label{eq:Kgaugefix}
\eeq
to obtain
\begin{equation}
-H_{\rm TC} = K \sum_p \prod_{l \in \partial p} \sigma^z_l + \Gamma_M \sum_s
\prod_{l: s \in \partial l} \sigma^x_l\ .
\end{equation}
which now acts in the subspace spanned solely by the link degrees of freedom.

One can see that the model is special by noting that all terms commute
with all others. This allows the full spectrum to be readily obtained
and one sees that the model exhibits topological order.
This is most easily done before gauge fixing, (Eq.~\ref{eq:Kgaugefix}), in
the basis of eigenstates of $\{\sigma^x_l \}$ and $\{\tau_s^x \}$.
In this formulation, sites where $\tau_s^x=-1$ are occupied by ``charges''
and links where $\sigma^x_l=-1$ carry ``electric flux'': $G_s=1$ is
the $Z_2$ version of Gauss's law. The dual, magnetic, flux passing
through loops $C$ is measured by the Wilson loops
$W[C] = \prod_{l \in C} \sigma^z_l$.

The ground states of $H_{\rm TC}$ lack charges, $\tau_s^x=1, \ \forall s$,
and are characterized by flux expulsion,
$\prod_{l \in \partial p} \sigma^z_l =1, \  \forall p$.

The complete lack of gauge field fluctuations, diagnosed by the Wilson loop $W = 1$, signals a deconfined phase.
Consistent with this ground state signature, the excited states
contain non-interacting charges, free to sit at arbitrary locations.

\subsection{Mapping to classical $Z_2$ gauge theory and $T>0$ phase transition}

We now turn to a computation of the partition function of Kitaev's
model at $T>0$. As gauge and matter sectors of the theory only interact via
the constraint, $G_s\equiv 1$, one can use  the
basis states $|\{\sigma_l \} \rangle \otimes |\{\tau_s \} \rangle$
where $\sigma^z_{l'} |\{\sigma_l \} \rangle = \sigma_{l'} |\{\sigma_l \}\rangle$
and $\tau^x_{s'} |\{\tau_s \} \rangle = \tau_{s'} |\{\tau_s \}\rangle$
to obtain
\beq
Z = Z_L \cdot Z_S
\eeq
where
\beq
Z_L = \sum_{\sigma_l = \pm 1} e^{\beta K \sum_p \prod_{l \in \partial p} \sigma_l}
\eeq
is the partition function of the classical or discretized Euclidean lattice $Z_2$ gauge theory in
$d$ dimensions and
\beq
Z_S = \sum_{\tau_s = \pm 1}  \delta_{\Pi_s \tau_s,1}
  e^{\beta \Gamma_m \sum_s \tau_s}
\eeq
is the partition function of independent spins in a magnetic field $\Gamma_m$ up to a single,
global, constraint that the number of flipped spins (in gauge language: charges) be even.
All the singularities in $Z$ thus are contained entirely in
$Z_L$.

The finite temperature phase structure of $Z_L$ is well known. In $d \ge 3$ it exhibits two phases. The
deconfined phase, at large $\beta K$, is characterized by a perimeter law for the Wilson
loop at large loop sizes. The confined phase, at small $\beta K$, exhibits an
area law. By contrast, in $d=2$, $Z_L$ exhibits only a confined phase, but with a confinement scale
(area law coefficient) $A_0$ that diverges exponentially as $\beta K \rightarrow \infty$.

All of this ties in with our previous results on Ising gauge theories:
(i) Kitaev's model exhibits a $T > 0$ continuation of the
$T=0$ topological phase in $d \ge 3$ but not in $d=2 $. (ii) the
presence of the topological phase can be detected by the (spatial) Wilson loop
exhibiting a perimeter law.   

\subsection{Stability of topological order under perturbations of the toric code}

It is often convenient to do calculations and point-of-principle demonstrations for topological phases at a ``Kitaev point'', where terms generically present in the Hamiltonian are set to vanish by hand, leaving behind a soluble model. This has some obvious drawbacks---for instance, a ``zero-law'' for the Wilson loop, $W \equiv 1 = e^{-0 \cdot L}$ is obviously somewhat non-generic, like a vanishing rather than finite correlation length in a disordered system, and any result thus obtained needs to be supplemented at least by a discussion of its fate under perturbations.

We now show that the deconfined phase
is characterized by a persistence of the $2^{N_{nc}}$-fold topological degeneracy that signals topological order, where $N_{nc}$ is the number of independent non-contractible loops. More precisely, it exhibits a cluster of states with splitting
of $O(e^{-L})$ for a system of linear dimension $L$ which are separated
from all other states by a gap of $O(L^0)$. To this end, we consider adding an
arbitrary local perturbation $V$ to the toric code, where locality now implies that the range
of variables coupled is strictly bounded, say by a distance $b$. For example, 
we may reintroduce the two terms in Eq.~\ref{IGTHamil} which we had set 
to zero on our way from the $Z_2$ gauge theory to the toric code, although
the argument to follow is completely general.

The easy part is to notice that the topologically distinct, degenerate ground states of $H_0$ will
not mix up to an
order $L/b \sim O(L)$ since they differ by the creation of a vortex loop (or appropriate
dimensional analog) that has to be created and stretched across the system in as many
applications of the perturbation. The less obvious part is that the states remain degenerate
in energy up to that order.

To see this, observe that the energy shifts $\delta E$ for two ground states, $| \psi_0 \rangle = | G \rangle$ (Eq. \ref{eq:GGs}) and $| \psi_1 \rangle = \Phi^x | \psi_0 \rangle$ differing by the flux through a non-contractible loop, $C_{\rm nc}$,
are sums of terms of the form
\begin{eqnarray}
\delta E_0 & = & \left< \psi_0 \left| VPV \dots \right| \psi_0 \right> \\
\delta E_1 & = & \left< \psi_0 \left| \Phi^x VPV \dots \Phi^x \right| \psi_0 \right> ~ .
\end{eqnarray}
Here $V$ is the local perturbation, $P = \left( 1- \sum_i | \psi_i \rangle \langle \psi_i | \right) \frac{1}{E_0 - H_0} \left( 1 - \sum_i | \psi_i \rangle \langle \psi_i | \right)$ projects onto the excited state manifold and $\Phi^x = \prod_{\ell \in C_{nc}} \sigma^x_\ell$ inserts a flux by operating on all links crossed by the non-contractible loop, $C_{nc}$.

Now,  we claim that $\delta E_0=\delta E_1$ to $o(L)$ in perturbation theory. To see this, note that we can permute one $\Phi^x$ through the operator products of $V$s and $P$s to annihilate the other. This is possible a) because $\Phi^x$ commutes with both numerator and denominator of $Pøââ$ and b) because any topologically equivalent deformation of the non-contractible loop along which the $\sigma^x$s act produces the same
state. Thus at an order below $L/b \sim O(L)$ in perturbation theory, there is always a choice of loop which avoids the locations of all the perturbing terms.

\end{document}